\documentstyle[revtex,preprint]{aps}
\begin{document}
\def\setstretch#1{\renewcommand{\baselinestretch}{#1}}
\setstretch{2.0}
\def\spacing#1{\par%
 \def\baselinestretch{#1}%
 \ifx\@currsize\normalsize\@normalsize\else\@currsize\fi}
\def\cl{\centerline}
\def\J{ J }
\def\ea                               {
     \epsilon_1                           }
\def\e                               {
     \epsilon                           }
\def\fl {f_{\sss L}(\e)}
\def\fr {f_{\sss R}(\e)}
\def\flw {f_{\sss L}(\w)}
\def\frw {f_{\sss R}(\w)}
\def\flr {f_{\sss L/R}}
\def\mulr {\mu_{\sss L/R}}
\def\sp                                {
     \hskip 0.02in                            }
\def\ss                               {
     \scriptstyle                       }
\def\sss                               {
     \scriptscriptstyle                       }
\def\raise                             {
     {\sss +}                            }
\def\k                               {
     k                            }
\def\w                               {
     \omega                             }
\def\c                               {
     {\bf c}                            }
\def\spin                            {
     \sigma                             }
\def\spinbar                         {
     {\bar \spin}                  }
\def\cspin                               {
     {\bf c}_{\spin}                   }
\def\cspindag                          {
     {\bf c}_{\spin}^{\raise}               }
\def\cbar                                {
     {\bf c}_{\spinbar}                   }
\def\cbardag                          {
     {\bf c}_{\spinbar}^{\raise}               }
\def\ckspin                               {
     {\bf c}_{\k\spin}                   }
\def\ckspindag                            {
     {\bf c}_{\k\spin}^{\raise}              }
\def\Vkspin                               {
     V_{\k\spin}                   }
\def\Vkbar                            {
     V_{\k\spinbar}                   }
\def\espin                           {
     \epsilon_{\spin}                   }
\def\ebar                        {
     \epsilon_{\spinbar}                }
\def\ekspin                           {
     \epsilon_{\k\spin}                   }
\def\ekbar                        {
     \epsilon_{\k\spinbar}                }
\def\sumk                            {
     \sum_{\k{\sss \in L,R} }                     }
\def\sumspin                         {
     \sum_{\spin}               }
\def\sumkspin                         {
     \sum_{\spin;\k{\sss \in L,R} }               }
\def\sumklorr                         {
     \sum_{\k {\sss \in L(R)} }               }
\def\nspin                            {
     n_{\spin}                         }
\def\nup                              {
     n_{\sss\uparrow}                         }
\def\ndown                            {
     n_{\sss\downarrow}                         }
\def\nbar                             {
     \langle n_{\spinbar} \rangle              }
\def\nspin                            {
     \langle n_{\spin} \rangle              }
\def\Gt                               {
     G_{\spin}(t)                         }
\def\Grt {G_{\spin}^{\,r}(t)}
\def\Gkt                               {
     G_{\k\spin}(t)                         }
\def\Glesst        {
     G_{\spin}^<(t)                         }
\def\Gtwo                              {
     G_{\spinbar \spin}(t)                         }
\def\Ge                               {
     G_{\spin}(\e)                         }
\def\Grw                               {
     G^{\,r}_{\spin}(\w)                         }
\def\Glessw        {
     G_{\spin}^<(\w)                         }
\def\Fw                               {
     F_{\spin}(\w)                         }
\def\fbar   {\bar f_{\spin}(\w)}
\def\sigzero                          {
     \Sigma_{{\sss 0}\spin}              }
\def\sigone                          {
     \Sigma_{{\sss 1}\spin}              }
\def\cond                             {
     \sigma                               }
\def\Gamalr                            {
     \Gamma^{\sss L(R)}_{\spin} (\w)              }
\def\Gama                            {
     \Gamma_{\spin} (\w)              }
\def\Gamal                            {
     \Gamma^{\sss L}_{\spin} (\w)              }
\def\Gamar                            {
     \Gamma^{\sss R}_{\spin} (\w)              }
\def\Gamblr                            {
     \Gamma^{\sss L(R)}_{{\sss 1}\spin} (\w)              }
\def\Gambl                            {
     \Gamma^{\sss L}_{{\sss 1}\spin} (\w)              }
\def\Gambr                            {
     \Gamma^{\sss R}_{{\sss 1}\spin} (\w)              }
\def\lifet {{1\over{\tau_{\sss \spin 1}}}   }
\def\ffd                            {
     f_{\sss FD}                               }
\def\eup                                {
    \epsilon_{\sss \uparrow}             }
\def\edown                              {
    \epsilon_{\sss \downarrow}             }
\def\kbt                                {
    k_{\!\sss B}T                            }
\def\tk                                 {
    T_{\!\sss K}                            }
\def\deltae                             {
    \Delta \e                              }
\def\eoms{equations-of-motion }
\def\eom{equations-of-motion}
\def\nca{non-crossing approximation}
\def\ncas{non-crossing approximation }
\def\doss{density of states }
\def\dos{density of states}
\def\lts{lifetime }
\def\lt{lifetime}
\def\squote{}
\def\quote#1#2#3#4{\squote {#1,\ {\sl#2}\ {\bf#3}, #4}.\par}
\def\qquote#1#2#3#4{\squote {#1,\ {\sl#2}\ {\bf#3}, #4};}
\def\nquote#1#2#3#4{\squote {#1,\ {\sl#2}\ {\bf#3}, #4}}
\def\book#1#2#3{\squote { #1,\ in {\sl#2}, edited by #3}.\par}
\def\nbook#1#2#3{\squote { #1,\ in {\sl#2}, edited by #3}}
\def\bbook#1#2#3{\squote { #1,\ in {\sl#2}, edited by #3}.}
\def\trans#1#2#3{[ {\sl #1} {\bf #2},\ #3 ]}
\def\prl{Phys. Rev. Lett.}
\def\pr{Phys. Rev.}

\newcount\rnumber
\def\defqoute#1{\newcount#1 #1=\rnumber\global\advance\rnumber by 1}
\rnumber=1
\def\fulde{1}
\def\anderson{2}
\def\reed{3}
\def\paul{4}
\def\leo{5}
\def\ashoori{6}
\def\meirav{7}
\def\mwl{8}
\def\beenakker{9}
\def\averin{10}
\def\glazman{11}
\def\ng{12}
\def\selman{13}
\def\bickers{14}
\def\lacroix{15}
\def\gunnarson{16}
\def\abrikosov{17}
\def\appelbaum{18}
\def\finiteu{19}
\def\mw{20}
\def\cox{21}
\def\pt{22}

\def\rpoint                          {     $====>$                        }
\def\lpoint                          {     $<====$                        }

\hsize 15 truecm
\hoffset -1.5 truecm
\setstretch{1.0}
{\Large \bf
\centerline{Low Temperature Transport through a Quantum Dot:}
\centerline{The Anderson Model out of Equilibrium}}
\bigskip
\author{Y. Meir}
\cl{Department of Physics, University of California}
\cl{Santa Barbara, CA 93106}
\medskip
\author{Ned S. Wingreen}
\cl{NEC Research Institute, 4 Independence Way, Princeton, NJ 08540}
\medskip
\author{Patrick A. Lee}
\cl{Department of Physics, Massachusetts Institute of Technology}
\cl{Cambridge, MA 02139}
\bigskip
\centerline{\sl abstract}
\setstretch{0.1}
\noindent{
The infinite-$U$ Anderson model is applied to non-equilibrium
transport through a quantum dot containing two spin levels weakly
coupled to two leads. At low temperatures,
the Kondo peak in the equilibrium \doss
 is split upon
the application of a voltage bias. The split peaks, one
at the chemical potential of each lead, are suppressed by
 non-equilibrium dissipation.
 In a magnetic field, the Kondo peaks shift
away from the chemical potentials by the Zeeman energy, leading to
an observable peak in the differential conductance when  the
non-equilibrium bias equals the Zeeman energy.
\setstretch{2.0}

\medskip

\bigskip

\newpage
The behavior of an atomic impurity coupled to conduction electrons
has become one of the paradigms of condensed matter
physics. Competition between on-site Coulomb interaction and
band hybridization produces the Kondo effect: a crossover from
weak to strong coupling between the localized and band electrons
below the Kondo temperature, $\tk$. The study of the Kondo effect has
been limited, however, by the nature of the impurity system.
Since it is a daunting task to drive the host metal out of equilibrium,
it is the {\it equilibrium} properties of Kondo impurities that have
been explored.$^{\fulde}$

In this paper we address a new Kondo system in which
{\it non-equilibrium} is routinely achieved, namely a
semiconductor quantum dot weakly coupled to its leads.
It is already evident that
Anderson's model$^{\anderson}$
 for a Kondo impurity - discrete, interacting levels
coupled to a band - also describes quantum
dots. Experimentally, the discrete spectrum of a single dot has been
probed by transport$^{\reed-\leo}$ and capacitance$^\ashoori$
spectroscopy, while the strong on-site Coulomb interaction is
observed in Coulomb-blockade conductance
oscillations.$^{\paul,\leo,\meirav}$ Theoretically, Anderson's model
has provided an
excellent description of these experiments
both in equilibrium,$^{\mwl,\beenakker}$ and
non-equilibrium.$^{\averin}$ However, it is only the high
temperature regime that has been explored experimentally,
while it is below $\tk$ that the Kondo effect emerges.

Since the Anderson Hamiltonian describes the quantum dot,
at low temperatures the dot must behave as a Kondo impurity.
In fact, Glazman \& Raikh$^{\glazman}$ and
Ng \& Lee$^{\ng}$ have
argued that at zero-temperature equilibrium
the Kondo resonance in the density of states of
spin-degenerate levels will always lead to perfect transparency
of the quantum dot at the Fermi energy.  In contrast,
above the Kondo temperature,
resonant tunneling occurs only at a discrete set of
Fermi energies. Thus the
Kondo effect will have a striking experimental signature in
low-temperature transport through a quantum dot.
Furthermore, in the quantum dot system the leads
coupled to the dot are easily biased to non-equilibrium
and the dot potential can be
swept continuously with a gate.
Thus new physical questions which were not relevant
to magnetic impurities can be raised. In particular, what
happens to the Kondo effect out of equilibrium ?$^{\selman}$
Since transport measurements on single quantum dots require
significant applied bias, this question is of immediate importance.

In this Letter we combine several approaches (\nca,$^{\bickers}$
 equations of motion,$^{\lacroix}$ perturbation theory,
variational wavefunction calculation$^{\gunnarson}$)
to present a consistent picture of low-temperature,
non-equilibrium transport through a quantum dot.
For spin-degenerate levels at equilibrium, the Kondo peak$^{\abrikosov}$
in the \doss at the chemical potential (Fig. 1a)  leads to
resonant transmission through the dot.$^{\glazman,\ng}$
A voltage bias between the left and right leads
causes the Kondo peak to split, leaving a peak in the \doss
at the chemical potential of each lead (Fig. 1b).
We find that the amplitudes of the split Kondo peaks
are suppressed by a finite  non-equilibrium lifetime.
This lifetime results from dissipative transitions
in which electrons are transferred from the high chemical potential lead
to the low chemical potential one. As the voltage bias is increased, the
lifetime decreases, resulting in
a
 suppression of the Kondo peaks and
thus a
 suppression of the differential conductance (Fig. 2a).$^{\selman}$
Upon application of a magnetic field, the Kondo peaks shift away from
the chemical potentials by the Zeeman splitting, but in opposite
directions for each spin (Figs. 1c and 1d). Interestingly, therefore, when the
chemical potential splitting equals the Zeeman splitting, a Kondo
peak shifted away from one chemical potential crosses the other
chemical potential. We predict an observable peak in the
differential conductance at this crossing$^{\appelbaum}$ (Fig. 2b).

We model the quantum dot and its leads by the
Anderson Hamiltonian$^{\anderson}$
\begin{equation}
 H = \sumkspin\!\! \ekspin \ckspindag \ckspin
  + \sumspin \espin \cspindag \cspin
  + U \nup \ndown + \sumkspin \!
       (\Vkspin \ckspindag \cspin + h.c.),\label{eq:H}
\end{equation}
where $\ckspindag (\ckspin)$  creates (destroys) an electron with momentum $k$
and spin $\spin$ in one of
the two leads, and $\cspindag (\cspin)$  creates (destroys) a
spin-$\spin$ electron on the quantum dot.
Since we are interested in temperatures
smaller than the orbital
level spacing in the quantum dot, we consider only
a single pair of levels on the dot with
energies $\eup = \e_0 + \deltae/2$ and  $\edown = \e_0 - \deltae/2$.
 The third term in (\ref{eq:H}) describes the Coulomb
interaction between the two localized spins
which we take to forbid double occupancy$^{\finiteu}$
($U \rightarrow \infty$),
while the fourth term
describes the hopping between the leads and the dot.

Our aim is to calculate the current through the dot, $J$, which for
 the case of proportionate couplings to the leads,
$\Gamal = \lambda \Gamar$, where
$ \Gamalr  = 2\pi\! \sumklorr |\Vkspin|^2 \,
   \delta(\w - \ekspin)$, 
can be expressed$^{\mw}$ in terms of the density of states,
 $-{1\over\pi} {\rm Im}\, \Grw$, as
\begin{equation}
\J = {e\over \hbar} \sum_{\spin} \int d\w \left[ \flw - \frw \right]
\,\Gama\, \Bigl[ -{1\over\pi} {\rm Im}\, \Grw  \Bigr] .
 \label {eq:J}
\end{equation}
In Eq. (\ref{eq:J}),
$\Gama = \Gamal\Gamar/\left[\Gamal+\Gamar\right]$, and  $\Grw$
     is the Fourier transform of the retarded Green function,
$ \Grt = -i\Theta(t) \langle \{ \cspin(t),\cspindag(0) \}
   \rangle. $

In order to calculate the Green function $\Grw$ we use both
the \nca$^{\bickers}$\ and an \eoms method.$^{\mwl,\lacroix,}$
The \ncas is based on an exact mapping
of the infinite-$U$ Anderson Hamiltonian (\ref{eq:H})
onto a slave-boson Hamiltonian. If vertex corrections are neglected,
the propagators for the boson and the
fermion degrees of freedom, which correspond, respectively,
to the propagators for
the empty site and a singly occupied site, obey a set of coupled
integral equations. Numerical solution of these equations
has been very useful in obtaining quantitative results for the
equilibrium system,$^{\bickers}$ including
the occupations of the two spin-states in the presence of
a magnetic field.$^{\cox}$ In this work we have generalized the \ncas to
non-equilibrium to produce densities of states, occupations, and
the nonlinear current (\ref{eq:J}).
However, as a large
spin-degeneracy (large $N$) technique, the \ncas produces a Kondo peak
even
 for the non-interacting
system ($N = 1$). Consequently, for $N = 2$ in a magnetic field,
it give rise to  spurious peaks in the density of states at
the chemical potentials.$^{\pt}$
Therefore, an \eoms method was  employed to complement the \ncas
and isolate its shortcomings. This method  corresponds to
a resummation of low-order hopping processes and cannot
produce a quantitative description of the Kondo effect.
Nevertheless, this method is known$^{\lacroix}$ to give the right qualitative
behavior at
 low temperatures.
 More importantly in the present context, since the equations-of-motion
method is exact for  $N=1$, it gives rise only
 to the proper Kondo peaks
(as identified by perturbation theory$^{\pt}$).

The \eoms method consists of differentiating the Green function $\Grt$ with
 respect to time, thereby generating higher-order Green functions which
 eventually have to be approximated to close the equation
 for $\Grt$. The procedure we employ here is the same as the one used
 in Ref. {\mwl}, which in the infinite-$U$ limit leads to
\begin{equation}
\Grw = { {1 - \nbar} \over {\w - \espin - \sigzero(\w) - \sigone(\w)}} \,,
    \label{eq:gfreq}
\end{equation}
with
\begin{equation}
\sigzero(\w) = \sumk {{|\Vkspin|^2 } \over {\w - \ekspin + i\eta} }\,,
\label{eq:selfen0}
\end{equation}
and
\begin{equation}
\sigone(\w) = \sumk {{|\Vkbar|^2 \flr(\ekbar)} \over
 {\w - \espin + \ebar - \ekbar + i\hbar/2 \tau_{\spinbar}} }\,,
\label{eq:selfen1}
\end{equation}
where $\flr(\e)$ is the Fermi distribution in the left/right lead
and  $\tau_{\spinbar}$ is the intermediate-state lifetime.
$\Grw$ has an overall amplitude proportional to
 $1-\nbar$, where $\nbar$ is the occupation  of the other spin-state.
Quantitative
 calculation of the occupations is beyond the scope of the equations of
motion in the present
 approximation scheme. Accordingly, we use the occupations resulting from
 the \nca, which are known to be quantitatively reliable in
equilibrium.$^{\cox}$

Within the \eoms scheme, the Kondo peak for spin $\sigma$ results from the
self-energy,
$\sigone(\w)$, due to virtual intermediate states in which the site is
occupied by an electron of opposite spin, $\spinbar$.
The remaining self-energy, $\sigzero(\w)$, is the exact self-energy
for the non-interacting case. Because of the sharp Fermi surfaces
at low temperature,
${\rm Re}\,\{ \sigone(\w)\}$ grows
logarithmically at $\w=\mu_{\sss L,R} \pm \deltae$, giving rise
 to peaks
in the density of states, $-{1\over\pi} {\rm Im}\, \Grw$,
near those energies.  The peaks
for the high-lying spin (low-lying spin)
appear near $\w=\mu_{\sss L,R}+ \deltae$
($\w=\mu_{\sss L,R}- \deltae$).
At zero-field and zero-temperature equilibrium,
the intermediate states giving rise
to $\sigone(\w)$ have an
infinite \lt, and the true peak in the \doss has an
amplitude corresponding to the unitarity limit.$^{\abrikosov}$ Once either
a voltage bias or a magnetic field is applied these
intermediate states acquire a finite
\lt, $\tau_{\spinbar}$,  which cuts off
the logarithmic divergence of ${\rm Re}\,\{\sigone(\w)\}$, resulting
in a suppression of the peak amplitudes. The \lt, $\tau_{\spin}$,
 of spin $\spin$ can
be calculated using second-order perturbation theory,
 yielding
\begin{eqnarray}
{1\over{\tau_{\spin}}}  =
 {1\over{\hbar}} \sum_{A=L,R} \Gamma^A_{\spin}(\espin)& \!\!\!\!
\left(1-f_A(\espin)\right)\, + \,
   \displaystyle{{{1\over4\pi\hbar}} \,  \sum_{{A,B=L,R}\atop{\sigma'}}}
\int_{-\infty}^{\infty} d\e  \, \,
 \left[{1 \over {(\espin-\e + i\eta)^2}}
  + {1 \over {(\espin-\e - i\eta)^2}} \right] \, \nonumber\\
& \times \left[ \Gamma^A_{\spin}(\e) \Gamma^B_{\spin'}(\e-\espin+\e_{\spin'})
 \left(1-f_A(\e)\right) f_B(\e-\espin+\e_{\spin'}) \right].
\label{eq:lifetime}
\end{eqnarray}
For a deep level at zero temperature and for constant $\Gamma$ this simplifies
to
\begin{equation}
{1\over{\tau_{\spin}}}  =
   \displaystyle{{{1\over2\pi\hbar}} \,  \sum_{{A,B=L,R}\atop{\sigma'}}}
\Gamma^A_{\spin}\,\Gamma^B_{\spin'}\,\Theta(\mu_B-\mu_A+\espin-\e_{\spin'})
{{\mu_B-\mu_A+\espin-\e_{\spin'}}\over{(\mu_A-\e_{\spin})
(\mu_B-\e_{\spin'})}},
\label{eq:lifet0}
\end{equation}
which explicitly shows that the \lts is non-zero only for finite bias or finite
magnetic field.

 In Fig. 1, we plot the density of states for  two spins symmetrically
coupled to two leads, consisting of Lorentzian bands of width
$2W$, so that $\Gamal=\Gamar=\Gamma W^2/2(\w^2+W^2)$, with $\Gamma\equiv 1$ and
$W=100$.
Results are shown for the \ncas (dashed lines), which is reliable
for zero magnetic field, and for the equations-of-motion method
(continuous lines),
which has the correct Kondo peak energies for all magnetic fields.
 In equilibrium and zero magnetic field,
the \doss exhibits  a single peak at the Fermi level
 as expected$^{\abrikosov}$ (Fig. 1a). As the
 chemical potentials split, the Kondo peak also splits, giving rise to a
 suppressed Kondo peak at each chemical potential (Fig. 1b).
  Upon the application of a magnetic field, the densities
of states for the two spins
 become different and the Kondo peaks shift away from the chemical
 potentials by the Zeeman splitting ($\deltae=0.2$ in Figs. 1c and 1d).
 The peaks move up in energy
 for the high-lying spin (Fig. 1c)
 and down in energy for the low-lying one (Fig. 1d).

The main conclusion of Fig. 1 is the emergence of new energy scales,
 not present in equilibrium.
 The Kondo peak in the equilibrium
density of states splits out of equilibrium
to two peaks spaced by the
chemical potential difference $\Delta \mu$, and  suppressed from equilibrium by
the finite dissipative
lifetime $\tau_{\sigma}$.
 In Fig. 1(b), the lifetime broadening, $\hbar/\tau_{\spin}$,
 is about the same as the
temperature.
It is apparent, however, from Figs. 1 (c) and (d),
that neither the \ncas nor the equations of motion
quantitatively determine the Kondo peaks at finite magnetic field.
For this case we use the equations-of-motion result since it
provides a good estimate of the Kondo peak positions (by comparison
with perturbation theory).
To understand the shift of the Kondo peaks
with magnetic field, it is helpful to recall how
the peaks in the \doss derive from the eigenstates of the
system. At $T=0$, $\Grt$
 involves
transitions from the $N$-particle ground state to all possible $N+1$
or $N-1$ states. At $B=0$ the correlated ground state has a finite amplitude
to have an
empty site, and thus $\cspindag$ ($\cspin$) can generate
transitions from the $N$-particle ground state to the
ground state with one more (one less) electron.
Since, by definition, the ground state energies differ
by the chemical potential, the \doss
includes a Kondo peak at the chemical potential.  Within a variational
calculation,$^{\gunnarson}$ we find that at finite magnetic field
the ground state is polarized, and adding or removing an electron
produces no overlap with the new ground state. However, there
is a correlated excited state of opposite polarization which can be reached,
and which consequently gives rise to a peak in the \dos, shifted
by the difference in energy between
polarization states, {\it i.e.} the Zeeman energy.

The current follows immediately from the densities of states
(\ref{eq:J}). In particular, the zero-temperature current
is the integrated density of states between the two chemical
potentials, weighted by the coupling to the leads $\Gama$.
At zero magnetic field, therefore, the Kondo
peak at the Fermi energy gives rise to a linear-response conductance
of $2e^2/h$ for symmetric barriers,
corresponding to perfect resonant transmission
through the quantum dot.$^{\glazman,\ng}$
As the bias is increased the differential conductance falls
rapidly (Fig. 2a).$^{\selman}$
This occurs firstly because the differential conductance
due to a peak in the density of states must fall off once
$\Delta\mu$ exceeds the peak width,
and secondly because the decreasing dissipative lifetime
suppresses the peak amplitudes. Since the peaks in the density
of states persist until the temperature is roughly one-tenth
the coupling to the leads, $\Gamma$, the peak in the differential
conductance is observable well above the Kondo temperature, $\tk$
(Fig. 2a, continuous line).


In a finite magnetic field the Kondo
peaks are shifted away from the chemical potential so they
contribute very little to the conductance in linear response.
As the bias is increased, however, the current carrying
region between the chemical potentials grows, until
at $\Delta \mu = \deltae$, it reaches one Kondo peak
in the \doss of each spin (see inset of Fig. 2b).
In Fig. 2b, one therefore sees peaks in the differential conductance
at $\Delta \mu = \deltae$ (continuous line). In fact, by
comparison with the non-crossing approximation (Fig. 1),
we expect the equations of motion
to underestimate the full strength of these peaks.
Experimentally, observation of peaks in the differential conductance
at $\Delta \mu = \deltae$
would provide a ``smoking gun" for the presence of Kondo physics
in transport through a quantum dot.


In this work, we addressed the non-equilibrium behavior of
Anderson's model for a magnetic impurity. Experimentally, the model
describes low-temperature transport
through a quantum dot, where non-equilibrium is readily accessible.
 We have shown
that new energy scales emerge in non-equilibrium. Specifically, the
difference in chemical potentials $\Delta \mu$ and the inverse
dissipation time  $\hbar/\tau_{\spin}$ lead, respectively, to splitting and
suppression of the Kondo resonances in the \dos.
Our results have led to a novel experimental prediction ---
when the Zeeman splitting
of the spins, $\deltae$, equals the applied bias, $\Delta\mu$,
 there will be a peak in the
differential conductance, provided these energies are smaller
than the coupling to the leads, $\Gamma$, and smaller than the
depth of the levels, $\mu_{\sss L,R}-\e_{\spin}$.
Importantly, this signature of the
Kondo effect persists, for a wide range of parameters,
to temperatures $\sim \Gamma/20$, which may
be magnitudes larger than the
Kondo temperature.
 We hope that this work will
encourage further efforts, both experimental and theoretical, to
probe the non-equilibrium physics of interacting quantum
systems.

{\sl Acknowledgments}: We thank W. Kohn, T. K. Ng,  and N. Read
 for valuable discussions.
Work at UCSB was supported by NSF
grant no. NSF-DMR90-01502
 and by the
NSF Science and Technology Center for Quantized Electronic Structures,
Grant no. DMR 91-20007.
Work at MIT was supported
under Joint Services Electronic Program Contract No. DAAL 03-89-0001.
\newpage
\small
Figure Captions:\\
(1) \, Density of states
 for an Anderson impurity symmetrically
coupled to two leads with chemical potentials $\mu_L$ and $\mu_R (= 0)$ and
Lorentzian bandwidth $2W$,
 from the \eoms method
(continuous line) and the \ncas (dashed line).
The impurity has two spin states with energies $\eup$ and $\edown$ and
an on-site interaction $U\rightarrow\infty$.
All energies are in units of the total coupling to the leads, $\Gamma$. The
band width is
$W=100$ and the temperature is $T=0.005$, roughly a factor of two
lower than the magnetization Kondo temperature
[Ref.  ${\bickers}$].
 (a) The equilibrium  ($\mu_L=0$)
\doss  at zero magnetic field $\eup=\edown=-2.0$.
The \doss exhibits  a single peak at the Fermi level
[Ref. ${\abrikosov}$].
(b) The  non-equilibrium ($\mu_L=0.3$) \doss
 at zero magnetic field $\eup=\edown=-2.0$. There is
 a suppressed Kondo peak at each chemical potential.
(c),(d) The  non-equilibrium
 ($\mu_L=0.3$) \doss for  spin up (c) and spin down
 (d) at finite magnetic field $\eup= -1.9, \edown=-2.1$.
 The Kondo peaks shift away from the chemical
 potentials by the Zeeman splitting $\deltae=0.2$;
 the shift is up in energy
 for the up spin and down in energy for the down spin.
\\
\bigskip
(2) \, Differential conductance, $e\,dJ/d\Delta\mu$,  with $\mu_R=0$,
 vs. applied bias. (a) Zero magnetic field differential conductance
via the non-crossing approximation.  (b) Differential conductance
 at the finite magnetic field, $\deltae=0.2$,
used in Figs. 1 (c) and (d), via equations of motion.
As shown in the inset, when the
 chemical potential difference, $\Delta \mu$,
reaches the Zeeman splitting, $\deltae$, the
 Kondo peaks in the \doss enter the region between the
chemical potentials, giving rise to a peak in the differential
conductance.
\newpage
\leftline{{\sl references:}}
\noindent
\begin{enumerate}
\setstretch{1.0}
\item For a review see \quote {P. Fulde, J. Keller, and G. Zwicknagl}
{Sol. Stat. Phys.} {41}{ 1 (1988)}
\item \quote{P. W. Anderson}{\pr}{124}{41 (1961)}
\item \qquote{M. A. Reed et al.}{\prl}{60}{535 (1988)}
\quote {C. G. Smith et al.}{J. Phys.}{C 21}{L893 (1988)}
\item \qquote{P.L. McEuen et al.}{Phys. Rev. Lett.}{66}{1926
 (1991)} E. B Foxman et al., unpublished.
\item \quote{A. T. Johnson et al.}{\prl}{69}{1592 (1992)}
\item \quote{R. Ashoori et al.}{\prl}{68}{3088 (1992)}
\item \quote{U. Meirav, M. Kastner, and S. J. Wind}{\prl}{65}{771 (1990)}
\item \quote{Y. Meir, N. S. Wingreen, and P. A. Lee}{\prl}{66}{3048
(1991)}
\item \quote{C.W.J. Beenakker}{Phys. Rev.}{B 44}{1646 (1991)}
\item \nquote{D.V. Averin and A.N. Korotkov}{Zh. Eksp. Teor. Fiz.}{97}
{927 (1990)} \trans{Sov. Phys. JETP}{70}{937 (1990)}
\item \quote {L. I. Glazman and M. E. Raikh} {JETP lett.}{47}{452 (1988)}
\item \quote {T. K. Ng and P. A. Lee}{\prl}{61}{1768 (1988)}
\item S. Hershfield, J. H. Davies, and J.W. Wilkins [{\sl \prl} {\bf 67}, 3720
 (1991)] have used a small $U$ expansion to study the symmetric
  Anderson model out
 of equilibrium in zero magnetic field. At finite bias,
 in contrast to our results, they find a single peak in the
 \doss centered between the two chemical potentials.
\item For a review see \quote {N. E. Bickers}{Rev. Mod. Phys.}{59}{845 (1987)}
\item \qquote {J. A. Appelbaum and D. R. Penn}{Phys. Rev.}{188}{874 (1969)}
\quote{C. Lacroix}{J. Phys. F}{11}{2389 (1981)}
\item \nquote {O. Gunnarsson and K. Sch\"{o}nhammer}{\prl}{50}{604 (1983)} and
 {\sl \pr} {\bf B 28}, 4315 (1983).
\item \qquote {A. A. Abrikosov}{Physics}{2}{5 and 61 (1965)}
\book {H. Suhl} {Theory of Magnetism in Transition Metals, course 37}
 {W. Marshall (Academic Press, New York, 1967)}
\item \nquote {J. A. Appelabaum}{\pr}{154}{633 (1967)}, has obtained shifted
divergences in the local Green function for the Kondo $s-d$ model.
\item The choice of $U\rightarrow\infty$  is appropriate for quantum
dots, where $U \sim 1 {\rm meV}$ and $\Gamma \sim 1-10 \mu{\rm eV}$
[see ref. {\paul}]. In fact, one expects very similar results for
finite $U$. When $U$ decreases below $W$, the bandwidth in the leads, it just
replaces $W$ as the large energy cutoff in the model [see ref. {\mwl}].
\item \quote{Y. Meir and N. S. Wingreen}{\prl}{68}{2512 (1992)}
\item \quote {D. L. Cox}{\pr}{B 35}{4561 (1987)}
\item The positions of the Kondo peaks in the density of states can
be determined from perturbation theory in the hopping. At $o(V^4)$,
the densities of states diverge logarithmically at $\mulr \pm
\deltae$ (N. S. Wingreen and Y. Meir, unpublished).
\end{enumerate}

\end{document}